\documentclass{article}
\usepackage{graphicx,times}
\begin {document}
\pagenumbering{arabic}
\pagestyle {plain}
\frenchspacing
\parindent 1.0 cm
\parskip 0.6cm
\vspace*{1.0 cm}
\begin{flushleft}
{\bf\large
Energy levels, radiative rates,  and lifetimes  for transitions in W LVIII} \vspace{0.5 cm}
\\{\sf Kanti M. Aggarwal} and {\sf Francis P. Keenan}\\ \vspace*{0.3 cm} 
Astrophysics Research Centre, School of Mathematics and Physics, Queen's University Belfast,\\Belfast BT7 1NN,
Northern Ireland, UK.\\  \vspace*{0.1 cm}

\vspace*{0.2 cm}  
{\bf ABSTRACT} \\ 

Energy levels and radiative rates are reported for transitions in Cl-like W LVIII.  Configuration interaction (CI) has been included among 44 configurations (generating 4978 levels) over a wide energy range up to 363 Ryd, and  the general-purpose relativistic atomic structure package ({\sc grasp})  adopted for the calculations.  Since no other results of comparable complexity are available, calculations have also been performed with the  flexible atomic code ({\sc fac}), which help in assessing the accuracy of our results. Energies are  listed for the lowest 400 levels (with energies up to $\sim$ 98 Ryd), which mainly belong to the  3s$^2$3p$^5$,  3s3p$^6$, 3s$^2$3p$^4$3d,  3s$^2$3p$^3$3d$^2$, 3s3p$^4$3d$^2$, 3s$^2$3p$^2$3d$^3$,   and 3p$^6$3d  configurations, and radiative rates are provided for four types of transitions, i.e. E1, E2, M1, and M2.  Our energy levels are assessed to be accurate to better than  0.5\%, whereas  radiative rates (and lifetimes) should be  accurate to  better than 20\% for a majority of the strong transitions. 

\vspace*{0.1 cm}  
Received January 29, 2014; accepted June 11, 2014\\ 
------------------------------------------------------------------------------------------------------------------------------------------------ \\
\vspace*{0.2 cm}
{\bf Running Title}: {\em K. M. Aggarwal and F. P. Keenan / Atomic Data and Nuclear Data Tables xxx (2014) xxx-xxx}
\end{flushleft}
\newpage

\begin{flushleft}
{\bf Contents}

\begin{tabular}{llr}
1.   &  Introduction  .........................................                &   00 \\
2.   &  Energy levels  ........................................            &   00 \\
3.   &  Radiative rates  ....................................           &   00 \\
4.   &  Lifetimes  .........................................                      &   00 \\
5.   &  Conclusions  ....................................		        &   00 \\
     &  Acknowledgments  .............................                   &   00 \\
     &  References  ........................................		         &   00 \\

\end{tabular}
\end{flushleft}
\begin{flushleft}
   Explanation of Tables \\ \vspace*{0.2 cm}   
   Tables\\ \vspace*{0.2 cm} 

\begin{tabular}{rlr}
                                                                                                                     
 1. & Configurations and  levels of W LVIII.  .................................................         &    00 \\
 2. &  Energies (Ryd) for the lowest 400 levels of W LVIII and their  lifetimes (in s). .................................................        &    00 \\
 3. & Transition wavelengths ($\lambda_{ij}$ in ${\rm \AA}$), radiative rates (A$_{ji}$ in s$^{-1}$), oscillator strengths  .................................... &    00 \\
     & (f$_{ij}$, dimensionless), line strengths S (in atomic unit) for electric dipole (E1), and ............................................. &    00 \\
     & A$_{ji}$ for electric quadrupole (E2), magnetic dipole (M1), and magnetic quadrupole (M2) transitions in W LVIII. .....                   &    00 \\ 
\end{tabular}	
\end{flushleft}															      

\newpage
\begin{flushleft}
{\bf 1. Introduction}
\end{flushleft}

\parindent = 1 cm 

Recently  there have been several studies ( \cite{kbf} -- \cite{jgk}) of atomic parameters (mainly energy levels and  radiative decay rates) for tungsten (W) ions. Similarly,  laboratory measurements  have been made for emission lines of W ions   -- see for example, Utter et al. \cite{sbu} and Clementson et al. \cite{ll1}.  These data  have been  compiled by the NIST (National Institute of Standards and Technology) team \cite{nist}, and are  available at their website {\tt http://physics.nist.gov/PhysRefData/ASD/levels\_form.html}. It is important to have such results because W is a major constituent of fusion  reactor devices, and hence data are required for assessing  and controlling  the radiation loss. With the ongoing  ITER project, the need for atomic data for several W ions has become even greater. 

Unfortunately, the only calculation available in the literature for (Cl-like)  W LVIII is  by  Mohan et al.  \cite{mas}, who have recently reported results for energy levels, oscillator strengths (f- values), radiative rates (A- values),  line strengths (S- values), and lifetimes ($\tau$). However, they listed energies for only 31 levels of the 3s$^2$3p$^5$, 3s3p$^6$, and 3s$^2$3p$^4$3d  configurations, and A- values for only transitions from the two levels of the ground state, i.e.  (3s$^2$3p$^5$) $^2$P$^o_{3/2,1/2}$. This amount of data is too limited for the modelling of plasmas, as demonstrated in Fig. 2 and Table 6 of Del Zanna et al. \cite{del} for   Cl-like iron (Fe X). Level populations deduced using restricted and larger sets of atomic data may  differ by up to a factor of five. Therefore, in this paper we  provide energies (and lifetimes) for the lowest 400 levels of W LVIII, which mainly belong to the  3s$^2$3p$^5$,  3s3p$^6$, 3s$^2$3p$^4$3d,  3s$^2$3p$^3$3d$^2$, 3s3p$^4$3d$^2$, 3s$^2$3p$^2$3d$^3$,   and 3p$^6$3d configurations. Furthermore, we also list A-values for all transitions among these levels, and for four types, namely electric  dipole (E1), electric  quadrupole (E2),  magnetic dipole (M1), and magnetic quadrupole (M2).

For the calculations, we have adopted the multi-configuration Dirac-Fock (MCDF) code,  developed by Grant  et al.  \cite{grasp0}. It is a fully relativistic code,  based on the $jj$ coupling scheme. Further higher-order relativistic corrections arising from the Breit interaction and QED (quantum electrodynamics) effects have also been included.  However, this initial version  has undergone several revisions, known as  GRASP  (general-purpose relativistic atomic structure package) \cite{grasp}, GRASP2 \cite{grasp92}, and GRASP2K \cite{grasp2k}--\cite{grasp2kk}. The version adopted here has been revised by one of its authors (Dr. P. H. Norrington), and is referred to as GRASP0. It  is freely available at  {\tt http://web.am.qub.ac.uk/DARC/}, and has been successfully used by ourselves and others to calculate atomic data for a wide range of ions. Furthermore, GRASP0 provides comparable results with other versions, and has also been employed by Mohan et al.  \cite{mas}.

\begin{flushleft}
{\bf 2. Energy levels}
\end{flushleft}

Since W is a heavy element (Z = 74), the contributions of  various relativistic operators are (expectedly) important in the determination of energy levels and subsequently other  parameters.  However,  the inclusion of {\em configuration interaction} (CI) is equally important for several of its ions --  see for example, Fournier \cite{kbf}. Furthermore, the importance of CI for Cl-like ions has already been established  -- see for example,  Ti VI \cite{tivi}, Cr VIII \cite{cr8}, Fe X \cite{fex}, and Co XI  \cite{coxi}.  For this reason Mohan et al.  \cite{mas} included CI among 15 configurations, namely 3s$^2$3p$^5$,  3s3p$^6$, 3s$^2$3p$^4$3d, 3s$^2$3p$^4$4$\ell$, 3s3p$^5$3d, 3s$^2$3p$^3$3d$^2$, 3s3p$^4$3d$^2$, 3p$^6$3d,   3s$^2$3p$^2$3d$^3$,  3p$^5$3d$^2$,  3s$^2$3p$^3$4d$^2$, and   3s$^2$3p$^3$4f$^2$. However, inclusion of this limited CI is not fully sufficient for an accurate determination of energy levels and radiative rates (A- values), as recently discussed by us \cite{w58}. 

In our earlier work \cite{w58}, we included extensive CI  among 20 even (1--20) and  18 odd (25--42) configurations, listed in Table 1. These 38 configurations generate a total of 3749 levels and cover a wide range of energy up to 363 Ryd. The 15 configurations  included by Mohan et al.  \cite{mas} are also listed in Table 1 (called GRASP1). It is clear from the table that these authors omitted several important configurations, such as 3s3p$^3$3d$^3$,  3s$^2$3p3d$^4$, and 3s3p$^2$3d$^4$, which together generate 1716 levels in the 92--196 Ryd energy range. The inclusion of these configurations substantially increases the size of a calculation, especially given that the 15 configurations of Mohan et al. generate only 1163 levels. However, the energy range of these three configurations is  well below that of 3s$^2$3p$^4$4$\ell$ included by Mohan et al. Levels of these three and other  configurations  (listed in Table 1)  closely interact, intermix, and influence the energies of others, such as 3s$^2$3p$^2$3d$^3$ and 3s$^2$3p$^3$3d$^2$ for even and odd parity levels. It is possible that Mohan et al. excluded most of the relevant configurations because they were only interested in the 31 levels of the 3s$^2$3p$^5$, 3s3p$^6$, and 3s$^2$3p$^4$3d  configurations, for which discrepancies in energy levels and A- values are not very appreciable \cite{w58}. However,  two of the configurations included by  them, namely (3s$^2$3p$^3$) 4d$^2$ and 4f$^2$,  are  not important, because their levels lie in a very high energy range, i.e. $\ge$ 413 Ryd, and hence have little effect on the lower levels. Therefore, we have {\em excluded} these two configurations from our calculations, but will discuss their impact  later. However, we have included a further 4 even (3s$^2$3d$^5$, 3p$^2$3d$^5$, 3s3d$^6$, and 3d$^7$) and 2 odd (3s3p3d$^5$ and 3p3d$^6$) configurations, which generate 1229 levels in the 157--303 Ryd energy range. All the 44 configurations included in our calculations are listed in Table 1 under the column GRASP2. Finally, as in our earlier work and that of  Mohan et al.,  we have adopted the option of {\em extended average level} (EAL),  in which a weighted (proportional to 2$j$+1) trace of the Hamiltonian matrix is minimised. This  yields results comparable to other options, such as {\em average level} (AL) -- see for example,  Aggarwal  et al.  for several ions of Kr \cite{kr} and Xe \cite{xe}.  

Energies for the lowest 400 levels of W LVIII are listed in Table 2, which correspond to the GRASP2 calculations with 44 configurations (listed in Table 1) generating 4978 levels in total. These listed levels mostly belong to the 3s$^2$3p$^5$,  3s3p$^6$, 3s$^2$3p$^4$3d,  3s$^2$3p$^3$3d$^2$, 3s3p$^4$3d$^2$, 3s$^2$3p$^2$3d$^3$,   and 3p$^6$3d configurations, but  energies for all levels may be obtained electronically on request from one of the authors (KMA: K.Aggarwal@qub.ac.uk). Unfortunately, experimental energies for W LVIII  (see the NIST website) are limited to only a few levels and are not very accurate because most are determined by interpolation and extrapolation. Nevertheless,  detailed comparisons with the NIST listings and those of Mohan et al.  \cite{mas}  were undertaken in our earlier work \cite{w58} for the lowest 31 levels belonging to the 3s$^2$3p$^5$, 3s3p$^6$, and 3s$^2$3p$^4$3d  configurations. As noted earlier,  the Mohan et al.  results are higher by up to 0.07 Ryd for some of the levels, mainly because of the limited CI included by them. We also note that for these 31 levels there are no differences between the present calculations (with a slightly larger CI) and the earlier ones \cite{w58}, in both magnitude and ordering. For the levels of other configurations, listed in Table 2,  it is  difficult to fully assess the accuracy of our energy levels, because no other  theoretical or experimental results exist. Therefore, to make such an  assessment  we have employed the  {\em Flexible Atomic Code} ({\sc fac}) of Gu \cite{fac},  available from the website {\tt http://sprg.ssl.berkeley.edu/$\sim$mfgu/fac/}. This is also a fully relativistic code which provides  results of comparable accuracy, particularly for energy levels and A- values, as already shown for several other ions, see for example:  Aggarwal  et al. for Kr \cite{kr} and Xe \cite{xe} ions, and Aggarwal and Keenan for Ti ions (\cite{tivi} and \cite{tivii}--\cite{tix}), and W XL \cite{w40}. 

As with {\sc grasp}, we have  performed a series of calculations  with the {\sc fac} code with increasing amount of CI,  but focus only on three, namely (i) FAC1, which includes 5821 levels among 3*7, 3s$^2$3p$^4$ 4*1,  3s$^2$3p$^4$ 5*1, 3s3p$^5$ 4*1, 3s3p$^5$ 5*1, 3p$^6$ 4*1, 3p$^6$ 5*1, and 3s$^2$3p$^3$3d 4*1; (ii) FAC2;  which includes an additional 3339 levels of   3s$^2$3p$^4$ 6*1,  3s3p$^5$ 6*1, 3p$^6$ 6*1, 3s$^2$3p$^3$3d 5*1, and 3s$^2$3p$^3$3d 6*1;   and finally (iii) FAC3, which  includes a further 9299 levels of  3s$^2$3p$^3$ 4*2, 3s$^2$3p$^3$ 5*2, 3s$^2$3p$^3$ 4*1 5*1, 3p$^5$ 4*2, and 3p$^5$3d 4*1, i.e. 18,459 levels in total.   

For Cl-like ions, CI is very important \cite{tivi}--\cite{coxi}, and levels from different configurations mix strongly. Therefore, it is very difficult to assign a unique label for each level. Mixing coefficients for the 31 levels of the 3s$^2$3p$^5$, 3s3p$^6$, and 3s$^2$3p$^4$3d  configurations were provided in Table 2 of  \cite{w58} from the GRASP1 calculations.  In Table A we list the mixing coefficients for the lowest 50 levels, which include three levels of the 3s$^2$3p$^5$ and  3s3p$^6$  configurations, but only a few of  3s$^2$3p$^4$3d and 3s$^2$3p$^3$3d$^2$. Moreover, these coefficients are  from our GRASP2 calculations and hence differ from those of \cite{w58}, particularly because all coefficients with magnitude $\ge$  $|$0.20$|$ are listed here. 

Some of the levels are almost pure, such as 3s$^2$3p$^5$ $^2$P$^o_{3/2, 1/2}$, and some are unambiguously identifiable (because of their clear dominance), such as: 3s$^2$3p$^4$($^3$P)3d $^4$D$_{3/2,5/2,7/2}$, i.e. 2, 3 and 7. However, some of the levels, such as 3s$^2$3p$^4$($^3$P)3d $^4$P$_{1/2,3/2,5/2}$, are highly mixed, and can therefore easily interchange with others.  Unfortunately, a majority of the W LVIII levels are highly mixed, as shown in Table A. For this reason, the $LSJ$ designations provided in Table 2 are only for guidance and should not be taken as definitive. The only confirm values associated with the levels are those of $J^{\pi}$.

In Table B we compare our energies for the lowest 50 levels from the GRASP2 calculations with those from FAC. Results from all three FAC calculations (FAC1, FAC2, and FAC3) described above are included in this table. Differences between the GRASP2 and FAC energies are up to 0.2 Ryd ($\sim$0.5\%) for some of the levels, such as 14--21. However, for a majority of the levels the discrepancies, if any, are below 0.1 Ryd, which is highly satisfactory. In general, the orderings from the two independent codes are also the same, although there are a few instances where these differ slightly, such as for 10/11 and 27/28. More importantly, all calculations from FAC agree within 0.03 Ryd and the orderings are also nearly the same. This clearly indicates that the additional CI included in the FAC2 and FAC3 calculations is of no advantage, and that included in the GRASP2 and FAC1 calculations is fully sufficient for the accurate determination of energy levels. Finally, based on this comparison as well as the one shown in Table 3 of  \cite{w58}, we can confidently state that our energy levels listed in Table 2 are accurate to $\sim$ 0.2 Ryd, or equivalently to 0.5\%.

\begin{flushleft}
{\bf 3. Radiative rates}
\end{flushleft}

The absorption oscillator strength ($f_{ij}$), a dimensionless quantity,  and radiative rate A$_{ji}$ (in s$^{-1}$) for a transition $i \to j$ are related by the following expression:

\begin{equation}
f_{ij} = \frac{mc}{8{\pi}^2{e^2}}{\lambda^2_{ji}} \frac{{\omega}_j}{{\omega}_i}A_{ji}
 = 1.49 \times 10^{-16} \lambda^2_{ji} \frac{{\omega}_j}{{\omega}_i} A_{ji}
\end{equation}
where $m$ and $e$ are the electron mass and charge, respectively, $c$  the velocity of light,  $\lambda_{ji}$  the transition wavelength in $\rm \AA$, and $\omega_i$ and $\omega_j$  the statistical weights of the lower $i$ and upper $j$ levels, respectively.
Similarly, the oscillator strength $f_{ij}$ (dimensionless) and the line strength $S$ (in atomic units, 1 a.u. = 6.460$\times$10$^{-36}$ cm$^2$ esu$^2$) are related by the 
following standard equations:

\begin{flushleft}
for the electric dipole (E1) transitions: 
\end{flushleft} 
\begin{equation}
A_{ji} = \frac{2.0261\times{10^{18}}}{{{\omega}_j}\lambda^3_{ji}} S \hspace*{1.0 cm} {\rm and} \hspace*{1.0 cm} 
f_{ij} = \frac{303.75}{\lambda_{ji}\omega_i} S, \\
\end{equation}
\begin{flushleft}
for the magnetic dipole (M1) transitions:  
\end{flushleft}
\begin{equation}
A_{ji} = \frac{2.6974\times{10^{13}}}{{{\omega}_j}\lambda^3_{ji}} S \hspace*{1.0 cm} {\rm and} \hspace*{1.0 cm}
f_{ij} = \frac{4.044\times{10^{-3}}}{\lambda_{ji}\omega_i} S, \\
\end{equation}
\begin{flushleft}
for the electric quadrupole (E2) transitions: 
\end{flushleft}
\begin{equation}
A_{ji} = \frac{1.1199\times{10^{18}}}{{{\omega}_j}\lambda^5_{ji}} S \hspace*{1.0 cm} {\rm and} \hspace*{1.0 cm}
f_{ij} = \frac{167.89}{\lambda^3_{ji}\omega_i} S, 
\end{equation}

\begin{flushleft}
and for the magnetic quadrupole (M2) transitions: 
\end{flushleft}
\begin{equation}
A_{ji} = \frac{1.4910\times{10^{13}}}{{{\omega}_j}\lambda^5_{ji}} S \hspace*{1.0 cm} {\rm and} \hspace*{1.0 cm}
f_{ij} = \frac{2.236\times{10^{-3}}}{\lambda^3_{ji}\omega_i} S. \\
\end{equation}

In our calculations with the {\sc grasp} code the S- (and subsequently A- and f-)  values have been determined in both the length and velocity forms, i.e. the Babushkin and Coulomb gauges in the relativistic nomenclature. Since the velocity form is  considered to be comparatively  less accurate,  in Table 3 we present results in the length form alone.   Included in this table are the  transition wavelengths ($\lambda_{ij}$ in ${\rm \AA}$), radiative rates (A$_{ji}$ in s$^{-1}$), oscillator strengths ($f_{ij}$, dimensionless), and line strengths ($S$ in a.u.) for all 19,862 electric dipole (E1) transitions among the lowest 400 levels of W LVIII.  Also, in calculating the  above parameters we have used the Breit and QED-corrected theoretical energies listed in Table 2, where the {\em indices} used to represent the lower and upper levels of a transition are also defined. However, only A- values are included in Table 3 for the 30,058 electric quadrupole (E2), 40,771  magnetic dipole (M1), and 28,365 magnetic quadrupole (M2) transitions. Corresponding results for f- or S- values can be  easily obtained using Eqs. (1-5). 

To assess the accuracy of our radiative data, in Table C we compare the f- values from our GRASP2  calculations with those from FAC1 and FAC2. All transitions with I $\le$ 5 and J $\le$ 50 are included in the table. Also,  the ratio  of the velocity/length forms is  listed, because it gives an indication of the accuracy of the f- (or A-) values. However, we emphasise here although good agreement between the two forms is  desirable,  it is not a necessary condition for accuracy. This is because different sets of configurations may lead to good agreement between the two forms, but entirely different results in magnitude, mostly for the weaker (inter-combination)  transitions, but sometimes also for  allowed ones, which are comparatively  stable and larger in magnitude. Examples of the differences between the two forms can be seen in some of our earlier papers, such as \cite{kma}--\cite{fe15}. 

The velocity/length ratio is generally within 20\% of unity for most of the transitions listed in Table C. However, for some weak transitions (such as 1--11, 2--38, 3--16, and 5--44/46), the ratio is higher, up to factor of 1.7. Nevertheless, most such transitions have f $\sim$ 10$^{-5}$ (or less) and their magnitudes are highly variable with differing amount of CI. Generally, with increasing amount of CI, the f- values for strong transitions converge, but this is not necessarily the case for weaker ones. For the same reasons, differences between the f- values from the GRASP and FAC calculations are up to a factor of two for some weak transitions, such as 1--11 and 5--26. Otherwise, for most transitions listed in Table C  the f- values from the GRASP and FAC calculations agree within 20\%, which is highly satisfactory. Finally, as for the energy levels, f- values from FAC1 and FAC2 are comparable for most transitions, and therefore confirm, yet again, that the CI included in the GRASP2 and FAC1 calculations is sufficient to produce accurate radiative rates. 

Quinet \cite{pq} has calculated the energy interval of the 3s$^2$3p$^5$ $^2$P$^o_{3/2,1/2}$ levels (1 and 10) to be 25.5738 Ryd which compares well with our result of 25.5358 Ryd  -- see Table 2. Like us he adopted the same GRASP0 version of the code and also calculated A- value for the 1--10 (M1 and E2) forbidden transition to be 3.95$\times$10$^8$ s$^{-1}$, which fully agrees with our result of  3.93$\times$10$^8$ s$^{-1}$. However, this comparison is too restricted and in the absence of the availability of other similar data,  assessing the reliability of the present calcuations  is not a simple task, as large discrepancies (of even orders of magnitude) have been noted in the past  \cite{fst}. Nevertheless, based on the comparisons shown in Table C and our experience on a variety of ions, the  accuracy of our radiative data is estimated to be better than 20\%, for a majority of transitions, particularly the stronger ones.

\begin{flushleft}
{\bf 4. Lifetimes}
\end{flushleft}

The lifetime $\tau$ of a level $j$ is defined as follows:

\begin{equation}
{\tau}_j = \frac{1}{{\sum_{i}^{}} A_{ji}}.
\end{equation}
In Table 2 we list lifetimes for all 400 levels from our calculations with the {\sc grasp} code, which  include A- values from all types of transitions, i.e. E1, E2, M1, and M2. Unfortunately, there are no measurements available with which to compare the lifetimes. However, theoretical results by Mohan et al.   \cite{mas} are  available for 31 levels of the  the 3s$^2$3p$^5$, 3s3p$^6$, and 3s$^2$3p$^4$3d  configurations,  based on calculations with the same {\sc grasp} code. However, these results are in error for several levels, by up to four orders of magnitude, as discussed and demonstrated in our earlier paper \cite{w58}. 

\begin{flushleft}
{\bf 5. Conclusions}
\end{flushleft}

In this work, energy levels and radiative rates (for E1, E2, M1, and M2 transitions)  obtained with the {\sc grasp} code have been reported for the lowest 400 levels of W LVIII. To assess the accuracy  of our  data, similar calculations with differing amount of CI  have also been performed with the {\sc fac} code, as no other results are available in the literature.  Based on comparisons of several calculations with the two independent codes, our energy levels are assessed to be accurate to $\sim$ 0.2 Ryd ($\sim$ 0.5\%), whereas the accuracy for other atomic parameters, including lifetimes, is estimated to be $\sim$ 20\%. Finally, energies have been calculated for up to 18,459 levels and A- values for up to 9160 levels, and these results (in electronic form) can be obtained from K.Aggarwal@qub.ac.uk. 
 
\section*{Acknowledgment}
 KMA  is thankful to  AWE Aldermaston for financial support.   

\begin{flushleft} 
{\bf Appendix A. Supplementary data} 
\end{flushleft} 

Owing to space limitations, only part of Table 3 is presented here, the full table being made available as supplemental material in conjunction with the electronic
publication of this work. Supplementary data associated with this article can be found, in the online version, at doi:nn.nnnn/j.adt.2014.nn.nnn.

\begin{flushleft}

\end{flushleft}


\newpage
\clearpage

\begin{table*}
\begin{flushleft}
{\bf Table A.}   Mixing coefficients (MC) for the lowest 50 levels of  W LVIII.  Numbers outside and inside a bracket correspond to MC and the level, respectively. See Table 2 for definitions of levels up to 400. \\
\end{flushleft}
\begin{tabular}{rlllrrrrrrrr} \hline
\\
Index  & Configuration           & Level              &  Mixing coefficients     \\
\\ \hline
\\
    1 & 3s$^2$3p$^5$			  & $^2$P$^o_{3/2}$   &    1.00(  1)										  \\
    2 & 3s$^2$3p$^4$($^3$P)3d		  & $^4$D$  _{3/2}$   &   -0.59(  2)+0.44( 12)-0.28( 56)+0.16(137)+0.46( 45)-0.28( 40)-0.23(  6)		  \\
    3 & 3s$^2$3p$^4$($^3$P)3d		  & $^4$D$  _{5/2}$   &   -0.36( 30)+0.61(  3)-0.25(165)-0.18( 47)+0.31( 13)+0.45( 39)-0.31( 50)		  \\
    4 & 3s$^2$3p$^4$($^3$P)3d		  & $^4$P$  _{1/2}$   &    0.36( 22)-0.71(  4)-0.22(  9)+0.46( 58)-0.32( 41)					  \\
    5 & 3s$^2$3p$^4$($^3$P)3d		  & $^2$F$  _{7/2}$   &    0.54( 42)-0.25(  7)+0.57(  5)+0.53( 31)-0.19( 52)					  \\
    6 & 3s$^2$3p$^4$($^1$S)3d		  & $^2$D$  _{3/2}$   &    0.44( 27)+0.30( 12)+0.21( 56)+0.24(137)-0.15( 40)+0.76(  6)  			  \\
    7 & 3s$^2$3p$^4$($^3$P)3d		  & $^4$D$  _{7/2}$   &   -0.23( 42)+0.65(  7)+0.47(  5)+0.21( 31)+0.50( 52)					  \\
    8 & 3s$^2$3p$^4$($^3$P)3d		  & $^4$F$  _{9/2}$   &    0.82(  8)+0.56( 48)  								  \\
    9 & 3s$^2$3p$^4$($^3$P)3d		  & $^2$P$  _{1/2}$   &   -0.11( 19)+0.24( 22)+0.35(  4)-0.67(  9)+0.36( 58)+0.47( 41)  			  \\
   10 & 3s$^2$3p$^5$			  & $^2$P$^o_{1/2}$   &    0.99( 10)										  \\
   11 & 3s$^2$3p$^4$($^1$S)3d		  & $^2$D$  _{5/2}$   &    0.22( 30)+0.47(165)-0.40( 47)+0.15( 13)+0.13( 39)+0.24( 50)+0.68( 11)		  \\
   12 & 3s$^2$3p$^4$($^3$P)3d		  & $^4$P$  _{3/2}$   &    0.20( 27)-0.47( 12)-0.44( 56)+0.45(137)+0.36( 45)+0.43( 40)  			  \\
   13 & 3s$^2$3p$^4$($^3$P)3d		  & $^2$D$  _{5/2}$   &   -0.28( 30)-0.41(  3)+0.22(165)-0.14( 47)+0.52( 13)+0.35( 39)+0.35( 50)-0.41( 11)	  \\
   14 & 3s$^2$3p$^3$($^2$D)3d$^2$($^3$F)  & $^2$D$^o_{3/2}$   &    0.37( 67)+0.19(111)-0.11(357)+0.24(459)-0.29(103)-0.26( 14)+0.12( 18)-0.22(174)	  \\
      & 				  & 		      &   -0.13(146)-0.38(294)+0.36( 23)+0.20(398)-0.15( 36)-0.25( 97)+0.23( 34)		  \\
   15 & 3s$^2$3p$^3$($^2$P)3d$^2$($^3$F)  & $^4$D$^o_{1/2}$   &   -0.45( 71)-0.24(415)-0.26(113)-0.11( 85)-0.31(299)-0.20( 51)-0.15(131)-0.59( 15)	  \\
      & 				  & 		      &   -0.11( 24)-0.33(368)  								  \\
   16 & 3s$^2$3p$^3$($^4$S)3d$^2$($^1$D)  & $^4$D$^o_{5/2}$   &   -0.35( 93)-0.26(121)-0.22( 16)-0.36( 68)-0.25(141)+0.10(187)-0.20(154)-0.28( 78)+	  \\
      & 				  & 		      &    0.41(290)-0.19(351)+0.29( 25)-0.14(171)+0.27( 44)					  \\
   17 & 3s$^2$3p$^3$($^2$D)3d$^2$($^3$F)  & $^4$H$^o_{7/2}$   &    0.24( 94)+0.24(160)+0.21( 87)-0.12(352)+0.25(144)-0.37( 17)-0.17(122)-0.25(289)+	  \\
      & 				  & 		      &    0.37(108)-0.20( 21)+0.37(293)-0.16( 29)+0.36( 46)					  \\
   18 & 3s$^2$3p$^3$($^2$D)3d$^2$($^3$P)  & $^4$F$^o_{3/2}$   &    0.14( 67)+0.33( 81)+0.21(178)+0.13(357)+0.27(166)-0.25( 18)+0.17(482)-0.12( 91)	  \\
      & 				  & 		      &   -0.17(136)+0.18(153)-0.27( 57)+0.16( 23)+0.28(117)-0.33(322)+0.22(385)+0.25( 36)	  \\
      & 				  & 		      &   -0.11(142)+0.37( 55)  								  \\
   19 & 3s3p$^6$			  & $^2$S$  _{1/2}$   &    0.85( 19)-0.11( 22)+0.31(  4)-0.13(  9)+0.19( 58)-0.31( 41)  			  \\
   20 & 3s$^2$3p$^3$($^2$D)3d$^2$($^3$F)  & $^4$H$^o_{9/2}$   &   -0.43( 92)-0.23(116)-0.14( 35)+0.35( 20)-0.12(423)-0.21( 73)+0.24(176)-0.13(133)	  \\
      & 				  & 		      &   -0.55(349)+0.28( 49)-0.24(168)-0.11(127)+0.17(381)					  \\
   21 & 3s$^2$3p$^3$($^2$P)3d$^2$($^3$F)  & $^4$F$^o_{7/2}$   &   -0.34( 94)-0.15(160)+0.14( 87)-0.23(352)+0.11( 32)+0.18(144)-0.30( 70)-0.10( 53)+	  \\
      & 				  & 		      &    0.16(122)-0.16(163)+0.13( 99)-0.17(124)-0.19(289)-0.25(108)+0.39( 21)+0.10(293)+	  \\
      & 				  & 		      &    0.24(360)-0.31( 29)-0.11(374)+0.11(139)+0.26( 46)					  \\
   22 & 3s$^2$3p$^4$($^3$P)3d		  & $^4$D$  _{1/2}$   &   -0.15( 19)-0.89( 22)-0.21(  4)-0.31(  9)+0.22( 58)					  \\
   23 & 3s$^2$3p$^3$($^2$P)3d$^2$($^3$F)  & $^4$D$^o_{3/2}$   &   -0.30( 67)+0.29(111)+0.19( 81)-0.18(357)-0.27( 88)-0.21(151)-0.22(188)+0.18( 18)+	  \\
      & 				  & 		      &    0.18(482)-0.11(294)-0.47( 23)+0.30(398)-0.23(322)+0.13(385)-0.24( 36)-0.10( 97)	  \\
   24 & 3s$^2$3p$^3$($^2$P)3d$^2$($^3$P)  & $^2$S$^o_{1/2}$   &   -0.21( 71)+0.37(145)-0.15(434)+0.17(415)-0.14(113)-0.11(299)+0.14(126)-0.41(169)+	  \\
      & 				  & 		      &    0.15(196)+0.13( 51)+0.14(131)-0.26( 15)-0.38(118)+0.44( 24)+0.25(368)		  \\
   25 & 3s$^2$3p$^3$($^2$P)3d$^2$($^3$F)  & $^2$F$^o_{5/2}$   &    0.23( 93)-0.28(121)-0.18(478)-0.21( 98)+0.11(110)+0.13( 16)-0.27( 68)-0.28(409)+	  \\
      & 				  & 		      &    0.15(177)-0.19(187)-0.11( 89)+0.14(134)+0.10(154)-0.20( 78)+0.36(351)+0.40( 25)	  \\
      & 				  & 		      &   -0.12( 26)+0.25(359)+0.15( 37)-0.15( 44)						  \\
   26 & 3s$^2$3p$^3$($^2$P)3d$^2$($^3$P)  & $^4$D$^o_{5/2}$   &   -0.23( 93)-0.43( 98)-0.12( 33)+0.14(409)-0.11(177)+0.14(187)+0.24(416)-0.15( 89)	  \\
      & 				  & 		      &   -0.12(143)+0.28(134)+0.15(290)-0.27(351)-0.11( 25)-0.41( 26)+0.41(359)-0.12(378)	  \\
      & 				  & 		      &   -0.11( 44)										  \\
   27 & 3s$^2$3p$^4$($^3$P)3d		  & $^4$F$  _{3/2}$   &   -0.57( 27)+0.57(  2)+0.19( 12)+0.36(137)+0.31( 45)-0.27( 40)  			  \\
   28 & 3s$^2$3p$^3$($^2$D)3d$^2$($^1$G)  & $^2$I$^o_{11/2}$  &    0.32(140)+0.39(410)-0.19(101)-0.11(120)+0.23(105)-0.38( 28)-0.12(161)+0.44( 43)+	  \\
      & 				  & 		      &    0.54(353)										  \\
   29 & 3s$^2$3p$^3$($^2$P)3d$^2$($^3$P)  & $^4$D$^o_{7/2}$   &    0.24( 87)+0.28(352)+0.28( 32)-0.16(144)-0.23( 17)-0.20( 70)-0.18( 99)+0.19(124)	  \\
      & 				  & 		      &   -0.20(130)-0.21(149)+0.12(289)+0.12( 21)+0.31(293)+0.18(360)+0.37( 29)-0.29(374)+	  \\
      & 				  & 		      &    0.26(139)-0.18( 46)  								  \\
   30 & 3s$^2$3p$^4$($^3$P)3d		  & $^4$F$  _{5/2}$   &    0.70( 30)+0.38( 47)+0.27( 13)+0.46( 39)-0.24( 50)					  \\

 \\ \hline                                                                                            
\end{tabular}       
\end{table*}    


\begin{table*}                                                                            
\begin{tabular}{rlllrrrrrrrr} \hline
\\
Index  & Configuration           & Level              &  Mixing coefficients     \\
\\ \hline
\\
   31 & 3s$^2$3p$^4$($^1$D)3d		  & $^2$G$  _{7/2}$   &     0.36( 42)-0.14(  7)+0.41(  5)-0.77( 31)+0.28( 52)					  \\ 
   32 & 3s$^2$3p$^3$($^4$S)3d$^2$($^1$G)  & $^4$G$^o_{7/2}$   &     0.29( 94)-0.23(160)-0.10( 87)-0.11(352)+0.27( 32)+0.26( 17)+0.13( 70)-0.14( 53)	  \\ 
      & 			    	  & 		      &    -0.20(453)+0.21(163)-0.20(130)-0.21(149)-0.32( 21)+0.29(170)-0.29(293)+0.15(360)	  \\ 
      & 			    	  & 		      &    -0.14( 29)-0.28(374)+0.26(139)							  \\ 
   33 & 3s$^2$3p$^3$($^4$S)3d$^2$($^3$P)  & $^4$P$^o_{5/2}$   &     0.20( 93)-0.13(478)-0.33( 33)-0.13(110)-0.19( 16)-0.18(409)-0.11( 84)+0.33(416)+	  \\ 
      & 			    	  & 		      &     0.15( 89)-0.13(143)-0.12(147)-0.20(154)-0.14(158)-0.24(290)+0.10(351)+0.18( 54)	  \\ 
      & 			    	  & 		      &    -0.29( 26)-0.17(359)-0.36(378)-0.18( 37)-0.14(171)+0.26( 44) 			  \\ 
   34 & 3s$^2$3p$^3$($^2$P)3d$^2$($^1$D)  & $^2$P$^o_{3/2}$   &     0.17( 67)+0.18(111)+0.16( 81)-0.28(178)-0.27(459)-0.20(103)+0.12( 18)-0.18(482)	  \\ 
      & 			    	  & 		      &    -0.26( 91)+0.11(136)+0.13(153)+0.23(174)+0.19(146)-0.27(294)+0.15(398)-0.15(117)+	  \\ 
      & 			    	  & 		      &     0.31(385)+0.32(142)+0.25( 97)-0.29( 34)						  \\ 
   35 & 3s$^2$3p$^3$($^4$S)3d$^2$($^1$G)  & $^4$G$^o_{9/2}$   &     0.24(116)-0.42( 35)-0.22( 20)-0.11(102)-0.39(133)-0.24(156)+0.19(349)+0.14( 49)+	  \\ 
      & 			    	  & 		      &     0.25(168)-0.33(127)+0.51(381)							  \\ 
   36 & 3s$^2$3p$^3$($^2$P)3d$^2$($^3$P)  & $^2$D$^o_{3/2}$   &     0.28(111)-0.27( 81)+0.10(178)+0.26(357)-0.16(103)-0.19( 88)-0.11(151)-0.12(188)	  \\ 
      & 			    	  & 		      &    -0.31( 18)-0.16(482)+0.15( 91)-0.16(153)-0.24(294)-0.21( 23)+0.25(398)+0.11(117)+	  \\ 
      & 			    	  & 		      &     0.27(322)-0.24(385)+0.39( 36)							  \\ 
   37 & 3s$^2$3p$^3$($^2$P)3d$^2$($^1$G)  & $^2$F$^o_{5/2}$   &    -0.24(478)+0.13( 33)-0.39(110)+0.13( 16)-0.11( 68)-0.12(409)-0.15( 84)-0.14(177)	  \\ 
      & 			    	  & 		      &    -0.15(416)-0.19(104)-0.33(147)+0.13(154)-0.13( 78)+0.18( 25)+0.26( 54)+0.15( 26)+	  \\ 
      & 			    	  & 		      &     0.14(378)-0.52( 37)-0.17( 44)							  \\ 
   38 & 3s$^2$3p$^3$($^2$P)3d$^2$($^3$P)  & $^2$P$^o_{1/2}$   &    -0.15( 71)-0.29(145)-0.30(434)+0.21(415)-0.44(126)-0.10(196)+0.19( 51)+0.14(131)	  \\ 
      & 			    	  & 		      &    -0.18( 15)-0.29( 95)+0.22(118)+0.47( 38)+0.30(368)					  \\ 
   39 & 3s$^2$3p$^4$($^1$D)3d		  & $^2$F$  _{5/2}$   &     0.22( 30)+0.37(  3)-0.20(165)+0.11( 47)+0.58( 13)-0.53( 39)+0.37( 50)		  \\ 
   40 & 3s$^2$3p$^4$($^1$D)3d		  & $^2$P$  _{3/2}$   &    -0.24( 27)-0.15(  2)+0.47( 12)-0.10( 56)+0.40(137)-0.50( 45)+0.52( 40)		  \\ 
   41 & 3s$^2$3p$^4$($^1$D)3d		  & $^2$S$  _{1/2}$   &     0.43( 19)-0.47(  4)-0.23(  9)-0.51( 58)+0.52( 41)					  \\ 
   42 & 3s$^2$3p$^4$($^3$P)3d		  & $^4$F$  _{7/2}$   &     0.62( 42)+0.66(  7)-0.30(  5)-0.26( 52)						  \\ 
   43 & 3s$^2$3p$^3$($^2$P)3d$^2$($^3$F)  & $^4$G$^o_{11/2}$  &     0.42(140)-0.29(410)-0.25(101)+0.26(105)+0.33( 28)+0.13(161)+0.53( 43)-0.44(353)	  \\ 
   44 & 3s$^2$3p$^3$($^2$P)3d$^2$($^1$D)  & $^2$D$^o_{5/2}$   &     0.15( 93)+0.15(478)-0.19( 98)+0.20( 33)-0.19(110)-0.29( 16)+0.13( 84)+0.19(177)	  \\ 
      & 			    	  & 		      &    -0.12(416)-0.20( 89)+0.12(134)-0.11(104)-0.14(147)-0.27(154)-0.12(158)+0.20(351)-	  \\ 
      & 			    	  & 		      &     0.10( 25)-0.21( 54)+0.31(359)+0.21(378)-0.25( 37)-0.18(171)+0.35( 44)		  \\ 
   45 & 3s$^2$3p$^4$($^1$D)3d		  & $^2$D$  _{3/2}$   &    -0.15( 27)-0.36(  2)-0.45( 12)+0.21( 56)+0.54(137)-0.33( 45)-0.45( 40)		  \\ 
   46 & 3s$^2$3p$^3$($^2$P)3d$^2$($^1$D)  & $^2$F$^o_{7/2}$   &    -0.20(160)+0.32(352)-0.17( 32)+0.28(144)-0.13( 53)-0.12(453)-0.20( 99)+0.17(124)+	  \\ 
      & 			    	  & 		      &     0.13(130)+0.11(149)-0.30(289)-0.12(180)-0.13(108)+0.19(170)+0.19(360)+0.39( 29)+	  \\ 
      & 			    	  & 		      &     0.18(374)-0.14(139)+0.41( 46)							  \\ 
   47 & 3s$^2$3p$^4$($^3$P)3d		  & $^2$F$  _{5/2}$   &    -0.21( 30)+0.38(  3)+0.47(165)+0.66( 47)-0.16( 13)+0.17( 39)+0.31( 50)		  \\ 
   48 & 3s$^2$3p$^4$($^1$D)3d		  & $^2$G$  _{9/2}$   &     0.56(  8)-0.82( 48) 								  \\ 
   49 & 3s$^2$3p$^3$($^2$P)3d$^2$($^3$F)  & $^4$F$^o_{9/2}$   &    -0.32( 92)+0.30(116)+0.21( 35)-0.18( 20)-0.34(102)-0.21(423)+0.12( 73)+0.20(176)+	  \\ 
      & 			    	  & 		      &     0.18(133)+0.50( 49)+0.37(168)+0.15(127)-0.23(381)					  \\ 
   50 & 3s$^2$3p$^4$($^1$D)3d		  & $^2$D$  _{5/2}$   &    -0.15( 30)+0.52(165)+0.11( 47)+0.35( 13)-0.36( 39)-0.65( 50) 			  \\ 
 \\ \hline  											      
\end{tabular}

\end{table*}

                               
\begin{flushleft}                                                                               
                               
{\small
                                                                               
}                                                                                               
                               
\end{flushleft} 

\newpage
\clearpage

\begin{table*}
\begin{flushleft}
{\bf Table B.}  Comparison of excitation energies (in Ryd) for the lowest 50 levels of W LVIII.
\end{flushleft}
\begin{tabular}{rllrrrrrrrrrr} \hline
Index  & Configuration                       & Level              & GRASP    & FAC1    & FAC2  & FAC3   \\
 \hline
    1  &  3s$^2$3p$^5$  		   &  $^2$P$^o_{3/2}$	  & 	 0.0000 &  0.0000   &  0.0000  &  0.0000 \\
    2  &  3s$^2$3p$^4$($^3$P)3d 	   &  $^4$D$  _{3/2}$	  & 	17.4003 & 17.2992   & 17.2906  & 17.3139 \\
    3  &  3s$^2$3p$^4$($^3$P)3d 	   &  $^4$D$  _{5/2}$	  & 	17.8570 & 17.7552   & 17.7459  & 17.7689 \\
    4  &  3s$^2$3p$^4$($^3$P)3d 	   &  $^4$P$  _{1/2}$	  & 	17.9195 & 17.8172   & 17.8071  & 17.8298 \\
    5  &  3s$^2$3p$^4$($^3$P)3d 	   &  $^2$F$  _{7/2}$	  & 	18.1268 & 18.0239   & 18.0138  & 18.0365 \\
    6  &  3s$^2$3p$^4$($^1$S)3d 	   &  $^2$D$  _{3/2}$	  & 	19.5583 & 19.4600   & 19.4516  & 19.4739 \\
    7  &  3s$^2$3p$^4$($^3$P)3d 	   &  $^4$D$  _{7/2}$	  & 	23.3137 & 23.2209   & 23.2113  & 23.2348 \\
    8  &  3s$^2$3p$^4$($^3$P)3d 	   &  $^4$F$  _{9/2}$	  & 	23.4798 & 23.3872   & 23.3781  & 23.4016 \\
    9  &  3s$^2$3p$^4$($^3$P)3d 	   &  $^2$P$  _{1/2}$	  & 	23.9746 & 23.8802   & 23.8693  & 23.8917 \\
   10  &  3s$^2$3p$^5$  		   &  $^2$P$^o_{1/2}$	  & 	25.5358 & 25.5597   & 25.5589  & 25.5590 \\
   11  &  3s$^2$3p$^4$($^1$S)3d 	   &  $^2$D$  _{5/2}$	  & 	25.6223 & 25.5309   & 25.5192  & 25.5407 \\
   12  &  3s$^2$3p$^4$($^3$P)3d 	   &  $^4$P$  _{3/2}$	  & 	26.1043 & 26.0095   & 25.9940  & 26.0153 \\
   13  &  3s$^2$3p$^4$($^3$P)3d 	   &  $^2$D$  _{5/2}$	  & 	26.4164 & 26.3209   & 26.3099  & 26.3320 \\
   14  &  3s$^2$3p$^3$($^2$D)3d$^2$($^3$F) &  $^2$D$^o_{3/2}$	  & 	35.8916 & 35.6898   & 35.6963  & 35.7203 \\
   15  &  3s$^2$3p$^3$($^2$P)3d$^2$($^3$F) &  $^4$D$^o_{1/2}$	  & 	36.3640 & 36.1609   & 36.1671  & 36.1918 \\
   16  &  3s$^2$3p$^3$($^4$S)3d$^2$($^1$D) &  $^4$D$^o_{5/2}$	  & 	36.3425 & 36.1396   & 36.1459  & 36.1709 \\
   17  &  3s$^2$3p$^3$($^2$D)3d$^2$($^3$F) &  $^4$H$^o_{7/2}$	  & 	36.4889 & 36.2847   & 36.2907  & 36.3151 \\
   18  &  3s$^2$3p$^3$($^2$D)3d$^2$($^3$P) &  $^4$F$^o_{3/2}$	  & 	38.1564 & 37.9439   & 37.9506  & 37.9659 \\
   19  &  3s$^2$3p$^6$  		   &  $^2$S$  _{1/2}$	  & 	39.8703 & 39.7896   & 39.7914  & 39.8009 \\
   20  &  3s$^2$3p$^3$($^2$D)3d$^2$($^3$F) &  $^4$H$^o_{9/2}$	  & 	41.6184 & 41.4287   & 41.4343  & 41.4602 \\
   21  &  3s$^2$3p$^3$($^2$P)3d$^2$($^3$F) &  $^4$F$^o_{7/2}$	  & 	41.9172 & 41.7274   & 41.7330  & 41.7572 \\
   22  &  3s$^2$3p$^4$($^3$P)3d 	   &  $^4$D$  _{1/2}$	  & 	42.0830 & 42.0099   & 42.0042  & 42.0280 \\
   23  &  3s$^2$3p$^3$($^2$P)3d$^2$($^3$F) &  $^4$D$^o_{3/2}$	  & 	42.5731 & 42.3810   & 42.3864  & 42.4106 \\
   24  &  3s$^2$3p$^3$($^2$P)3d$^2$($^3$P) &  $^2$S$^o_{1/2}$	  & 	42.5963 & 42.4049   & 42.4108  & 42.4321 \\
   25  &  3s$^2$3p$^3$($^2$P)3d$^2$($^3$F) &  $^2$F$^o_{5/2}$	  & 	42.6822 & 42.4888   & 42.4938  & 42.5189 \\
   26  &  3s$^2$3p$^3$($^2$P)3d$^2$($^3$P) &  $^4$D$^o_{5/2}$	  & 	42.7646 & 42.5717   & 42.5780  & 42.5991 \\
   27  &  3s$^2$3p$^4$($^3$P)3d 	   &  $^4$F$  _{3/2}$	  & 	42.8703 & 42.7967   & 42.7886  & 42.8123 \\
   28  &  3s$^2$3p$^3$($^2$D)3d$^2$($^1$G) &  $^2$I$^o_{11/2}$    & 	42.8909 & 42.6992   & 42.7046  & 42.7295 \\
   29  &  3s$^2$3p$^3$($^2$P)3d$^2$($^3$P) &  $^4$D$^o_{7/2}$	  & 	43.1573 & 42.9633   & 42.9684  & 42.9918 \\
   30  &  3s$^2$3p$^4$($^3$P)3d 	   &  $^4$F$  _{5/2}$	  & 	43.4667 & 43.3925   & 43.3830  & 43.4062 \\
   31  &  3s$^2$3p$^4$($^1$D)3d 	   &  $^2$G$  _{7/2}$	  & 	43.5688 & 43.4938   & 43.4847  & 43.5082 \\
   32  &  3s$^2$3p$^3$($^4$S)3d$^2$($^1$G) &  $^4$G$^o_{7/2}$	  & 	44.0459 & 43.8484   & 43.8511  & 43.8763 \\
   33  &  3s$^2$3p$^3$($^4$S)3d$^2$($^3$P) &  $^4$P$^o_{5/2}$	  & 	44.1611 & 43.9626   & 43.9669  & 43.9890 \\
   34  &  3s$^2$3p$^3$($^2$P)3d$^2$($^1$D) &  $^2$P$^o_{3/2}$	  & 	44.3757 & 44.1760   & 44.1804  & 44.2022 \\
   35  &  3s$^2$3p$^3$($^4$S)3d$^2$($^1$G) &  $^4$G$^o_{9/2}$	  & 	44.6977 & 44.4984   & 44.5000  & 44.5246 \\
   36  &  3s$^2$3p$^3$($^2$P)3d$^2$($^3$P) &  $^2$D$^o_{3/2}$	  & 	45.0197 & 44.8173   & 44.8217  & 44.8434 \\
   37  &  3s$^2$3p$^3$($^2$P)3d$^2$($^1$G) &  $^2$F$^o_{5/2}$	  & 	45.5541 & 45.3527   & 45.3560  & 45.3801 \\
   38  &  3s$^2$3p$^3$($^2$P)3d$^2$($^3$P) &  $^2$P$^o_{1/2}$	  & 	45.7114 & 45.5099   & 45.5148  & 45.5356 \\
   39  &  3s$^2$3p$^4$($^1$D)3d 	   &  $^2$F$  _{5/2}$	  & 	45.9153 & 45.8408   & 45.8279  & 45.8495 \\
   40  &  3s$^2$3p$^4$($^1$D)3d 	   &  $^2$P$  _{3/2}$	  & 	46.0974 & 46.0249   & 46.0094  & 46.0309 \\
   41  &  3s$^2$3p$^4$($^1$D)3d 	   &  $^2$S$  _{1/2}$	  & 	46.6321 & 46.5486   & 46.5400  & 46.5585 \\
   42  &  3s$^2$3p$^4$($^3$P)3d 	   &  $^4$F$  _{7/2}$	  & 	48.3089 & 48.2449   & 48.2368  & 48.2609 \\
   43  &  3s$^2$3p$^3$($^2$P)3d$^2$($^3$F) &  $^4$G$^o_{11/2}$    & 	48.5175 & 48.3351   & 48.3385  & 48.3638 \\
   44  &  3s$^2$3p$^3$($^2$P)3d$^2$($^1$D) &  $^2$D$^o_{5/2}$	  & 	48.8295 & 48.6486   & 48.6529  & 48.6753 \\
   45  &  3s$^2$3p$^4$($^1$D)3d 	   &  $^2$D$  _{3/2}$	  & 	49.0045 & 48.9379   & 48.9261  & 48.9492 \\
   46  &  3s$^2$3p$^3$($^2$P)3d$^2$($^1$D) &  $^2$F$^o_{7/2}$	  & 	49.1876 & 49.0051   & 49.0093  & 49.1963 \\
   47  &  3s$^2$3p$^4$($^3$P)3d 	   &  $^2$F$  _{5/2}$	  & 	49.3072 & 49.2416   & 49.2297  & 49.2328 \\
   48  &  3s$^2$3p$^4$($^1$D)3d 	   &  $^2$G$  _{9/2}$	  & 	49.2839 & 49.2187   & 49.2091  & 49.0314 \\
   49  &  3s$^2$3p$^3$($^2$P)3d$^2$($^3$F) &  $^4$F$^o_{9/2}$	  & 	49.3527 & 49.1681   & 49.1706  & 49.2523 \\
   50  &  3s$^2$3p$^4$($^1$D)3d 	   &  $^2$D$  _{5/2}$	  & 	49.8203 & 49.7519   & 49.7384  & 49.7608 \\
 \hline            								                	 
\end{tabular}   								   					       
			      							   					       

														       
\begin{flushleft}													       
{\small
GRASP: present calculations from the {\sc grasp} code with 4978 levels\\ 
FAC1: present calculations from the {\sc fac} code with 5821 levels  \\  
FAC2: present calculations from the {\sc fac} code with 9160 levels \\ 
FAC3: present calculations from the {\sc fac} code with 18,459 levels \\																       															       
}															       
\end{flushleft} 
\end{table*}

\newpage
\clearpage

\begin{table*}
\begin{flushleft}
{\bf Table C.}  Comparison of oscillator strengths (f- values) for some transitions of W LVIII. See Table 2 for level indices.  $a{\pm}b \equiv a{\times}$10$^{{\pm}b}$.  \\
\end{flushleft}

\begin{tabular}{rrrrrrrrrrrr} \hline
     I &   J    & GRASP      &       FAC1 &       FAC2 & Ratio   &     I  &   J     & GRASP      &       FAC1 &       FAC2 & Ratio   \\
 \hline                                                                         
     1  &  2  &  1.0882-3  & 1.081-3 & 1.077-3  &   1.0-0   &	3 &  18  &  2.6720-3  & 2.639-3 & 2.636-3   &  9.7-1  \\
     1  &  3  &  2.4344-3  & 2.430-3 & 2.430-3  &   1.1-0   &	3 &  21  &  2.3411-5  & 2.132-5 & 2.138-5   &  8.7-1  \\
     1  &  4  &  3.8991-3  & 3.868-3 & 3.859-3  &   1.1-0   &	3 &  23  &  4.3433-3  & 4.284-3 & 4.277-3   &  1.0-0  \\
     1  &  6  &  5.9768-4  & 5.986-4 & 6.004-4  &   1.1-0   &	3 &  25  &  3.6256-3  & 3.617-3 & 3.611-3   &  9.8-1  \\
     1  &  9  &  9.0692-3  & 8.975-3 & 8.963-3  &   1.1-0   &	3 &  26  &  4.2007-4  & 3.789-4 & 3.689-4   &  1.0-0  \\
     1  & 11  &  4.4697-5  & 6.737-5 & 9.514-5  &   1.1-0   &	3 &  29  &  5.2679-4  & 5.322-4 & 5.282-4   &  1.1-0  \\
     1  & 12  &  9.3168-2  & 9.295-2 & 9.274-2  &   1.0-0   &	3 &  32  &  1.2444-1  & 1.242-1 & 1.238-1   &  1.1-0  \\
\hline                                                                                               
\end{tabular}                                                                                   
                            
                               
\begin{flushleft}                                                                               
                               
{\small
GRASP: present calculations from the {\sc grasp} code with 4978 levels\\ 
FAC1: present calculations from the {\sc fac} code with 5821 levels  \\  
FAC2: present calculations from the {\sc fac} code with 9160 levels \\    
Ratio: ratio of velocity and length forms of f- values from the {\sc grasp} code \\                                                                                 
}                                                                                               
                               
\end{flushleft} 

\end{table*}


\newpage
\clearpage

\begin{flushleft}
{\bf Explanation of Tables} \\  \vspace{0.2 cm}
\end{flushleft}

\begin{flushleft}
Table 1. Configurations and  levels of W LVIII.  \\ \vspace{0.2 cm}
\begin{tabular}{ll}
Index            &  Configuration Index \\
Configuration    & The configuration included in the calculation \\
No. of Levels             & The number of levels the configuration generates. Odd parity levels are designated with a superscript ``o" \\
Energy  Range         & Energy range of the levels in Ryd unit \\
GRASP1            &     Calculations of Mohan et al. \cite{mas} from the {\sc grasp} code with 1163 levels \\ 
GRASP2           &      Present calculations from the {\sc grasp} code with 4978 levels\\										
Y                          &     If a configuration is included under a calculation \\ 
\end{tabular}
\end{flushleft}

\begin{flushleft}
Table 2. Energies (Ryd) for the lowest 400 levels of W LVIII and their lifetimes ($\tau$, s).  \\ \vspace{0.2 cm}
\begin{tabular}{ll}
Index            & Level Index \\
Configuration    & The configuration to which the level belongs \\
Level             & The $LSJ$ designation of the level \\
Energy           & Present energies from the {\sc grasp} code  with 4978 level calculations \\
$\tau$ (s)       & Lifetime of the level in s \\
\end{tabular}
\end{flushleft}

\begin{flushleft}
Table 3. Transition wavelengths ($\lambda_{ij}$ in $\rm \AA$), radiative rates (A$_{ji}$ in s$^{-1}$),
 oscillator strengths (f$_{ij}$, dimensionless), and line strengths (S, in atomic units) for electric dipole (E1), and 
A$_{ji}$ for electric quadrupole (E2), magnetic dipole (M1), and magnetic quadrupole (M2) transitions of W LVIII. \\ \vspace{0.2 cm}
\begin{tabular}{ll}
$i$ and $j$         & The lower ($i$) and upper ($j$) levels of a transition as defined in Table 2.\\
$\lambda_{ij}$      & Transition wavelength (in ${\rm \AA}$) \\
A$^{E1}_{ji}$       & Radiative transition probability (in s$^{-1}$) for the E1 transitions \\
f$^{E1}_{ij}$       & Absorption oscillator strength (dimensionless) for the E1 transitions \\
S$^{E1}$            & Line strength in atomic unit (a.u.), 1 a.u. = 6.460$\times$10$^{-36}$ cm$^2$ esu$^2$ for the E1 transitions \\
A$^{E2}_{ji}$       & Radiative transition probability (in s$^{-1}$) for the E2 transitions \\
A$^{M1}_{ji}$       & Radiative transition probability (in s$^{-1}$) for the M1 transitions \\
A$^{M2}_{ji}$       & Radiative transition probability (in s$^{-1}$) for the M2 transitions \\
$a{\pm}b$ &  $\equiv a\times{10^{{\pm}b}}$ \\
\end{tabular}
\end{flushleft}


\newpage
\clearpage

\begin{flushleft}
Table 1. Configurations/Levels of W LVIII and their energy ranges (Ryd). 
\end{flushleft}
\begin{tabular}{rllcrrrrrrrrr} \hline
Index  & Configuration      & No. of Levels  & Energy Range   &  GRASP1 & GRASP2    \\
 \hline
i) even \\
  1  &  3s$^2$3p$^4$3d        &     28       & 17-75	  &  Y &  Y    \\
  2  &  3s3p$^6$	      &     1	     & 40	  &  Y &  Y    \\
  3  &  3s$^2$3p$^2$3d$^3$    &    261       & 55-130	  &  Y &  Y    \\
  4  &  3s3p$^4$3d$^2$        &    211	     & 73-143	  &  Y &  Y    \\
  5  &  3p$^6$3d	      &    2	     & 100-106    &  Y &  Y    \\
  6  &  3s3p$^2$3d$^4$        &    858       & 113-196    &    &  Y    \\
\hline  											      
\end{tabular}   								   					       
			      							   					       
\begin{flushleft}													       
{\small
GRASP1: earlier calculations of Mohan et al. \cite{mas} from the {\sc grasp} code with 1163 levels \\ 
GRASP2: present calculations from the {\sc grasp} code with 4978 levels\\										
Y: configuration included under a calculation \\ 										       
															       
}															       
\end{flushleft} 

\clearpage
\newpage

\begin{flushleft}
Table 2. Energies (Ryd) for the lowest 400 levels of W LVIII and their lifetimes ($\tau$, s). $a{\pm}b \equiv a{\times}$10$^{{\pm}b}$.
\end{flushleft}
\begin{tabular}{rllrrrrrrrrrr} \hline
Index  & Configuration                       & Level              & Energy  & $\tau$  (s)   \\
 \hline			       
    1  &  3s$^2$3p$^5$  		   &  $^2$P$^o_{3/2}$	  & 	 0.0000 & ........  \\
    2  &  3s$^2$3p$^4$($^3$P)3d 	   &  $^4$D$  _{3/2}$	  & 	17.4003 & 3.778-10  \\
    3  &  3s$^2$3p$^4$($^3$P)3d 	   &  $^4$D$  _{5/2}$	  & 	17.8570 & 2.406-10  \\
    4  &  3s$^2$3p$^4$($^3$P)3d 	   &  $^4$P$  _{1/2}$	  & 	17.9195 & 4.972-11  \\
    5  &  3s$^2$3p$^4$($^3$P)3d 	   &  $^2$F$  _{7/2}$	  & 	18.1268 & 4.557-03  \\
    6  &  3s$^2$3p$^4$($^1$S)3d 	   &  $^2$D$  _{3/2}$	  & 	19.5583 & 5.445-10  \\
    7  &  3s$^2$3p$^4$($^3$P)3d 	   &  $^4$D$  _{7/2}$	  & 	23.3137 & 4.054-07  \\
\hline            								                	 
\end{tabular}  

\clearpage
\newpage
\begin{tabular}{rllrrrrrrrrrr} \hline
Index  & Configuration                       & Level              & Energy  & $\tau$  (s)   \\
 \hline
   51  &  3s$^2$3p$^3$($^2$D)3d$^2$($^1$D)  & $^2$P$^o_{1/2}$	  & 	50.2590 & 1.476-12  \\
   52  &  3s$^2$3p$^4$($^1$D)3d 	    & $^2$F$  _{7/2}$	  & 	50.4435 & 3.433-09  \\
   53  &  3s$^2$3p$^3$($^2$D)3d$^2$($^3$F)  & $^4$F$^o_{7/2}$	  & 	51.4400 & 1.056-12  \\
   54  &  3s$^2$3p$^3$($^2$P)3d$^2$($^3$F)  & $^2$D$^o_{5/2}$	  & 	51.6981 & 1.569-12  \\
   55  &  3s$^2$3p$^3$($^2$P)3d$^2$($^1$S)  & $^2$P$^o_{3/2}$	  & 	51.7675 & 2.541-12  \\
   56  &  3s$^2$3p$^4$($^3$P)3d 	    & $^2$D$  _{3/2}$	  & 	52.1902 & 1.185-12  \\
   57  &  3s$^2$3p$^3$($^2$D)3d$^2$($^1$S)  & $^2$D$^o_{3/2}$	  & 	52.6962 & 8.744-13  \\
\hline            								                	 
\end{tabular}  

\clearpage
\newpage
\begin{tabular}{rllrrrrrrrrrr} \hline
Index  & Configuration                       & Level              & Energy  & $\tau$  (s)   \\
\hline
  101  &  3s$^2$3p$^3$($^2$D)3d$^2$($^3$F) &  $^4$H$^o_{11/2}$    & 	66.6766 & 1.886-10  \\
  102  &  3s$^2$3p$^3$($^2$D)3d$^2$($^3$F) &  $^4$G$^o_{9/2}$	  & 	66.7611 & 5.194-12  \\
  103  &  3s$^2$3p$^3$($^2$D)3d$^2$($^3$F) &  $^4$F$^o_{3/2}$	  & 	67.4468 & 1.103-12  \\
  104  &  3s$^2$3p$^3$($^2$D)3d$^2$($^1$G) &  $^2$F$^o_{5/2}$	  & 	67.4624 & 3.185-12  \\
  105  &  3s$^2$3p$^3$($^2$D)3d$^2$($^3$F) &  $^2$H$^o_{11/2}$    & 	67.4485 & 2.098-11  \\
  106  &  3s$^2$3p$^2$($^1$S)3d$^3$($^4$F) &  $^4$F$  _{7/2}$	  & 	68.1596 & 1.822-12  \\
  107  &  3s$^2$3p$^3$($^2$D)3d$^2$($^1$G) &  $^2$I$^o_{13/2}$    & 	67.9341 & 1.543-08  \\
\hline            								                	 
\end{tabular}  

\clearpage
\newpage
\begin{tabular}{rllrrrrrrrrrr} \hline
Index  & Configuration                       & Level              & Energy  & $\tau$  (s)   \\
\hline
  151  &  3s$^2$3p$^3$($^2$D)3d$^2$($^3$F) &  $^4$P$^o_{3/2}$	  & 	73.2832 & 3.040-13  \\
  152  &  3s$^2$3p$^3$($^2$D)3d$^2$($^1$D) &  $^2$G$^o_{9/2}$	  & 	73.1988 & 9.664-12  \\
  153  &  3s$^2$3p$^3$($^2$D)3d$^2$($^3$P) &  $^2$P$^o_{3/2}$	  & 	73.5774 & 1.803-13  \\
  154  &  3s$^2$3p$^3$($^2$D)3d$^2$($^1$D) &  $^2$F$^o_{5/2}$	  & 	73.5310 & 9.704-13  \\
  155  &  3s3p$^4$($^4$P)3d$^2$($^3$F)     &  $^6$D$  _{3/2}$	  & 	73.9922 & 2.100-12  \\
  156  &  3s$^2$3p$^3$($^2$D)3d$^2$($^1$G) &  $^2$G$^o_{9/2}$	  & 	73.8308 & 2.368-12  \\
  157  &  3s3p$^4$($^4$P)3d$^2$($^3$F)     &  $^6$F$  _{5/2}$	  & 	74.3251 & 1.753-12  \\
\hline            								                	 
\end{tabular}  

\clearpage
\newpage
\begin{tabular}{rllrrrrrrrrrr} \hline
Index  & Configuration                       & Level              & Energy  & $\tau$  (s)   \\
\hline
  201  &  3s3p$^4$($^4$P)3d$^2$($^3$F)     &  $^4$F$  _{3/2}$	  & 	81.3905 & 4.891-13  \\
  202  &  3s3p$^4$($^2$D)3d$^2$($^3$F)     &  $^2$D$  _{3/2}$	  & 	81.6634 & 3.439-13  \\
  203  &  3s$^2$3p$^2$($^3$P)3d$^3$($^4$F) &  $^6$G$  _{5/2}$	  & 	81.6945 & 3.348-13  \\
  204  &  3s3p$^4$($^2$D)3d$^2$($^3$F)     &  $^2$P$  _{1/2}$	  & 	81.8367 & 2.876-13  \\
  205  &  3s3p$^4$($^4$P)3d$^2$($^3$F)     &  $^4$F$  _{5/2}$	  & 	81.9080 & 1.354-12  \\
  206  &  3s3p$^4$($^4$P)3d$^2$($^3$F)     &  $^4$D$  _{3/2}$	  & 	82.0205 & 9.677-13  \\
  207  &  3s$^2$3p$^2$($^1$D)3d$^3$($^2$G) &  $^2$G$  _{7/2}$	  & 	82.0228 & 3.176-13  \\
\hline            								                	 
\end{tabular}  

\clearpage
\newpage
\begin{tabular}{rllrrrrrrrrrr} \hline
Index  & Configuration                       & Level              & Energy  & $\tau$  (s)   \\
\hline
  251  &  3s3p$^4$($^2$P)3d$^2$($^1$G)      & $^2$G$  _{7/2}$	  & 	86.3352 & 4.327-13  \\
  252  &  3s3p$^4$($^4$P)3d$^2$($^1$D)      & $^4$F$  _{9/2}$	  & 	86.3991 & 1.673-12  \\
  253  &  3s$^2$3p$^2$($^1$D)3d$^3$($^4$F)  & $^4$D$  _{1/2}$	  & 	86.7377 & 4.524-13  \\
  254  &  3s$^2$3p$^2$($^3$P)3d$^3$($^4$F)  & $^4$F$  _{3/2}$	  & 	86.7530 & 4.255-13  \\
  255  &  3s$^2$3p$^2$($^1$D)3d$^3$($^2$G)  & $^2$I$  _{13/2}$    & 	86.6163 & 1.328-12  \\
  256  &  3s$^2$3p$^2$($^1$D)3d$^3$($^2$G)  & $^2$D$  _{5/2}$	  & 	86.8589 & 4.673-13  \\
  257  &  3s$^2$3p$^2$($^3$P)3d$^3$($^2$G)  & $^4$G$  _{7/2}$	  & 	86.8310 & 5.673-13  \\
\hline            								                	 
\end{tabular}  

\clearpage
\newpage
\begin{tabular}{rllrrrrrrrrrr} \hline
Index  & Configuration                       & Level              & Energy  & $\tau$  (s)   \\
 \hline
  301  &  3s3p$^5$($^3$P)3d		   &  $^4$D$^o_{3/2}$	  & 	90.8633 & 3.011-13  \\
  302  &  3s3p$^4$($^2$S)3d$^2$($^3$F)     &  $^2$F$  _{7/2}$	  & 	90.8591 & 6.707-13  \\
  303  &  3s3p$^5$($^3$P)3d		   &  $^2$F$^o_{5/2}$	  & 	90.8305 & 3.239-13  \\
  304  &  3s3p$^4$($^2$S)3d$^2$($^3$P)     &  $^2$P$  _{3/2}$	  & 	91.1510 & 5.605-13  \\
  305  &  3s$^2$3p$^2$($^1$D)3d$^3$($^2$G) &  $^2$F$  _{7/2}$	  & 	91.2677 & 2.399-13  \\
  306  &  3s$^2$3p$^2$($^3$P)3d$^3$($^2$D) &  $^4$P$  _{5/2}$	  & 	91.3920 & 1.935-13  \\
  307  &  3s$^2$3p$^2$($^1$D)3d$^3$($^2$F) &  $^2$G$  _{7/2}$	  & 	91.3596 & 5.674-13  \\
\hline            								                	 
\end{tabular}  

\clearpage
\newpage
\begin{tabular}{rllrrrrrrrrrr} \hline
Index  & Configuration                       & Level              & Energy  & $\tau$  (s)   \\
\hline
  351  &  3s$^2$3p$^3$($^2$P)3d$^2$($^3$F) &  $^4$D$^o_{5/2}$	  & 	94.5885 & 2.107-13  \\
  352  &  3s$^2$3p$^3$($^4$S)3d$^2$($^3$P) &  $^6$P$^o_{7/2}$	  & 	94.5717 & 2.698-13  \\
  353  &  3s$^2$3p$^3$($^2$P)3d$^2$($^1$G) &  $^2$H$^o_{11/2}$    & 	94.6042 & 5.138-13  \\
  354  &  3s$^2$3p$^2$($^3$P)3d$^3$($^2$D) &  $^4$F$  _{5/2}$	  & 	94.8229 & 1.151-12  \\
  355  &  3s$^2$3p$^2$($^3$P)3d$^3$($^4$F) &  $^4$D$  _{7/2}$	  & 	94.8309 & 4.948-13  \\
  356  &  3s$^2$3p$^2$($^3$P)3d$^3$($^2$H) &  $^4$H$  _{13/2}$    & 	94.7132 & 3.702-13  \\
  357  &  3s$^2$3p$^3$($^4$S)3d$^2$($^3$P) &  $^2$P$^o_{3/2}$	  & 	94.9166 & 2.582-13  \\
\hline            								                	 
\end{tabular}   								   					       
			      							   					       

\begin{flushleft}													       
{\small										       
															       
}															       
\end{flushleft} 


\clearpage
\newpage

\setcounter{table}{2}

\begin{table*}                                                                                                                                                  
\caption{Transition wavelengths ($\lambda_{ij}$ in ${\rm \AA}$), radiative rates (A$_{ji}$ in s$^{-1}$), oscillator strengths (f$_{ij}$, dimensionless), and line     
strengths (S, in atomic units) for electric dipole (E1), and A$_{ji}$ for E2, M1, and M2 transitions in W LVIII. ($a{\pm}b \equiv a{\times}$10$^{{\pm}b}$).}        
\begin{tabular}{rrrrrrrrr}                                                                                                                                      
\hline                                                                                                                                                          
\hline                                                                                                                                                          
$i$ & $j$ & $\lambda_{ij}$ & A$^{{\rm E1}}_{ji}$  & f$^{{\rm E1}}_{ij}$ & S$^{{\rm E1}}$ & A$^{{\rm E2}}_{ji}$  & A$^{{\rm M1}}_{ji}$ & A$^{{\rm M2}}_{ji}$ \\  
\hline                                                                                                                                                      
    1 &    2 &  5.237$+$01 &  2.647$+$09 &  1.088$-$03 &  7.505$-$04 &  0.000$+$00 &  0.000$+$00 &  3.430$+$00 \\       
    1 &    3 &  5.103$+$01 &  4.157$+$09 &  2.434$-$03 &  1.636$-$03 &  0.000$+$00 &  0.000$+$00 &  1.466$+$01 \\       
    1 &    4 &  5.085$+$01 &  2.011$+$10 &  3.899$-$03 &  2.611$-$03 &  0.000$+$00 &  0.000$+$00 &  3.636$+$01 \\       
    1 &    5 &  5.027$+$01 &  0.000$+$00 &  0.000$+$00 &  0.000$+$00 &  0.000$+$00 &  0.000$+$00 &  3.598$+$01 \\       
    1 &    6 &  4.659$+$01 &  1.836$+$09 &  5.977$-$04 &  3.667$-$04 &  0.000$+$00 &  0.000$+$00 &  1.470$+$02 \\       
    1 &    7 &  3.909$+$01 &  0.000$+$00 &  0.000$+$00 &  0.000$+$00 &  0.000$+$00 &  0.000$+$00 &  3.608$+$04 \\       
    1 &    9 &  3.801$+$01 &  8.374$+$10 &  9.069$-$03 &  4.539$-$03 &  0.000$+$00 &  0.000$+$00 &  9.836$+$04 \\
\hline                                                                                                                                                          
\end{tabular}                                                                                                                                                   
\end{table*}                                                                                                                                                    
\newpage                                                                                                                                                        
\clearpage                                                                                                                                                      
\setcounter{table}{2}                                                                                                                                           
\begin{table*}                                                                                                                                                  
\caption{Transition wavelengths ($\lambda_{ij}$ in ${\rm \AA}$), radiative rates (A$_{ji}$ in s$^{-1}$), oscillator strengths (f$_{ij}$, dimensionless), and line     
strengths (S, in atomic units) for electric dipole (E1), and A$_{ji}$ for E2, M1, and M2 transitions in W LVIII. ($a{\pm}b \equiv a{\times}$10$^{{\pm}b}$).}    
\begin{tabular}{rrrrrrrrr}                                                                                                                                      
\hline                                                                                                                                                          
\hline                                                                                                                                                          
$i$ & $j$ & $\lambda_{ij}$ & A$^{{\rm E1}}_{ji}$  & f$^{{\rm E1}}_{ij}$ & S$^{{\rm E1}}$ & A$^{{\rm E2}}_{ji}$  & A$^{{\rm M1}}_{ji}$ & A$^{{\rm M2}}_{ji}$ \\  
\hline                                                                                                                                                          
    1 &   59 &  1.640$+$01 &  2.522$+$08 &  1.017$-$05 &  2.195$-$06 &  0.000$+$00 &  0.000$+$00 &  2.296$+$01 \\       
    1 &   60 &  1.632$+$01 &  0.000$+$00 &  0.000$+$00 &  0.000$+$00 &  1.017$+$07 &  1.952$+$03 &  0.000$+$00 \\       
    1 &   61 &  1.598$+$01 &  0.000$+$00 &  0.000$+$00 &  0.000$+$00 &  2.893$+$07 &  3.837$+$03 &  0.000$+$00 \\       
    1 &   62 &  1.582$+$01 &  0.000$+$00 &  0.000$+$00 &  0.000$+$00 &  1.280$+$08 &  0.000$+$00 &  0.000$+$00 \\       
    1 &   63 &  1.578$+$01 &  0.000$+$00 &  0.000$+$00 &  0.000$+$00 &  5.520$+$07 &  4.160$+$04 &  0.000$+$00 \\       
    1 &   64 &  1.553$+$01 &  0.000$+$00 &  0.000$+$00 &  0.000$+$00 &  1.099$+$08 &  1.317$+$06 &  0.000$+$00 \\       
    1 &   65 &  1.540$+$01 &  0.000$+$00 &  0.000$+$00 &  0.000$+$00 &  5.818$+$07 &  1.734$+$03 &  0.000$+$00 \\
\hline                                                                                                                                                          
\end{tabular}                                                                                                                                                   
\end{table*}                                                                                                                                                    
\newpage                                                                                                                                                        
\clearpage                                                                                                                                                      
\setcounter{table}{2}                                                                                                                                           
\begin{table*}                                                                                                                                                  
\caption{Transition wavelengths ($\lambda_{ij}$ in ${\rm \AA}$), radiative rates (A$_{ji}$ in s$^{-1}$), oscillator strengths (f$_{ij}$, dimensionless), and line     
strengths (S, in atomic units) for electric dipole (E1), and A$_{ji}$ for E2, M1, and M2 transitions in W LVIII. ($a{\pm}b \equiv a{\times}$10$^{{\pm}b}$).}    
\begin{tabular}{rrrrrrrrr}                                                                                                                                      
\hline                                                                                                                                                          
\hline                                                                                                                                                          
$i$ & $j$ & $\lambda_{ij}$ & A$^{{\rm E1}}_{ji}$  & f$^{{\rm E1}}_{ij}$ & S$^{{\rm E1}}$ & A$^{{\rm E2}}_{ji}$  & A$^{{\rm M1}}_{ji}$ & A$^{{\rm M2}}_{ji}$ \\  
\hline                                                                                                                                                          
    1 &  121 &  1.312$+$01 &  0.000$+$00 &  0.000$+$00 &  0.000$+$00 &  2.888$+$06 &  2.309$+$03 &  0.000$+$00 \\       
    1 &  122 &  1.312$+$01 &  0.000$+$00 &  0.000$+$00 &  0.000$+$00 &  3.699$+$07 &  0.000$+$00 &  0.000$+$00 \\       
    1 &  124 &  1.302$+$01 &  0.000$+$00 &  0.000$+$00 &  0.000$+$00 &  2.512$+$07 &  0.000$+$00 &  0.000$+$00 \\       
    1 &  125 &  1.299$+$01 &  5.245$+$07 &  1.990$-$06 &  3.403$-$07 &  0.000$+$00 &  0.000$+$00 &  5.658$-$05 \\       
    1 &  126 &  1.298$+$01 &  0.000$+$00 &  0.000$+$00 &  0.000$+$00 &  2.466$+$07 &  1.712$+$04 &  0.000$+$00 \\       
    1 &  128 &  1.298$+$01 &  0.000$+$00 &  0.000$+$00 &  0.000$+$00 &  0.000$+$00 &  0.000$+$00 &  1.344$+$01 \\       
    1 &  129 &  1.297$+$01 &  0.000$+$00 &  0.000$+$00 &  0.000$+$00 &  7.440$+$04 &  1.133$+$04 &  0.000$+$00 \\
\hline                                                                                                                                                          
\end{tabular}                                                                                                                                                   
\end{table*}                                                                                                                                                    
\newpage                                                                                                                                                        
\clearpage                                                                                                                                                      
\setcounter{table}{2}                                                                                                                                           
\begin{table*}                                                                                                                                                  
\caption{Transition wavelengths ($\lambda_{ij}$ in ${\rm \AA}$), radiative rates (A$_{ji}$ in s$^{-1}$), oscillator strengths (f$_{ij}$, dimensionless), and line     
strengths (S, in atomic units) for electric dipole (E1), and A$_{ji}$ for E2, M1, and M2 transitions in W LVIII. ($a{\pm}b \equiv a{\times}$10$^{{\pm}b}$).}    
\begin{tabular}{rrrrrrrrr}                                                                                                                                      
\hline                                                                                                                                                          
\hline                                                                                                                                                          
$i$ & $j$ & $\lambda_{ij}$ & A$^{{\rm E1}}_{ji}$  & f$^{{\rm E1}}_{ij}$ & S$^{{\rm E1}}$ & A$^{{\rm E2}}_{ji}$  & A$^{{\rm M1}}_{ji}$ & A$^{{\rm M2}}_{ji}$ \\  
\hline                                                                                                                                                          
    1 &  183 &  1.171$+$01 &  6.671$+$08 &  1.371$-$05 &  2.114$-$06 &  0.000$+$00 &  0.000$+$00 &  5.396$+$02 \\       
    1 &  184 &  1.174$+$01 &  0.000$+$00 &  0.000$+$00 &  0.000$+$00 &  1.343$+$07 &  1.228$+$05 &  0.000$+$00 \\       
    1 &  185 &  1.169$+$01 &  9.837$+$08 &  3.024$-$05 &  4.655$-$06 &  0.000$+$00 &  0.000$+$00 &  8.775$+$01 \\       
    1 &  186 &  1.169$+$01 &  0.000$+$00 &  0.000$+$00 &  0.000$+$00 &  0.000$+$00 &  0.000$+$00 &  7.370$+$02 \\       
    1 &  187 &  1.171$+$01 &  0.000$+$00 &  0.000$+$00 &  0.000$+$00 &  1.143$+$07 &  1.215$+$04 &  0.000$+$00 \\       
    1 &  188 &  1.162$+$01 &  0.000$+$00 &  0.000$+$00 &  0.000$+$00 &  3.743$+$07 &  2.831$+$04 &  0.000$+$00 \\       
    1 &  189 &  1.148$+$01 &  5.389$+$08 &  5.320$-$06 &  8.040$-$07 &  0.000$+$00 &  0.000$+$00 &  9.710$+$02 \\
\hline                                                                                                                                                          
\end{tabular}                                                                                                                                                   
\end{table*}                                                                                                                                                    
\newpage                                                                                                                                                        
\clearpage                                                                                                                                                      
\setcounter{table}{2}                                                                                                                                           
\begin{table*}                                                                                                                                                  
\caption{Transition wavelengths ($\lambda_{ij}$ in ${\rm \AA}$), radiative rates (A$_{ji}$ in s$^{-1}$), oscillator strengths (f$_{ij}$, dimensionless), and line     
strengths (S, in atomic units) for electric dipole (E1), and A$_{ji}$ for E2, M1, and M2 transitions in W LVIII. ($a{\pm}b \equiv a{\times}$10$^{{\pm}b}$).}    
\begin{tabular}{rrrrrrrrr}                                                                                                                                      
\hline                                                                                                                                                          
\hline                                                                                                                                                          
$i$ & $j$ & $\lambda_{ij}$ & A$^{{\rm E1}}_{ji}$  & f$^{{\rm E1}}_{ij}$ & S$^{{\rm E1}}$ & A$^{{\rm E2}}_{ji}$  & A$^{{\rm M1}}_{ji}$ & A$^{{\rm M2}}_{ji}$ \\  
\hline                                                                                                                                                          
    1 &  245 &  1.059$+$01 &  0.000$+$00 &  0.000$+$00 &  0.000$+$00 &  7.150$+$04 &  1.634$+$06 &  0.000$+$00 \\       
    1 &  246 &  1.057$+$01 &  9.831$+$03 &  1.648$-$10 &  2.294$-$11 &  0.000$+$00 &  0.000$+$00 &  1.966$+$03 \\       
    1 &  247 &  1.057$+$01 &  2.819$+$07 &  2.360$-$07 &  3.285$-$08 &  0.000$+$00 &  0.000$+$00 &  7.562$+$01 \\       
    1 &  248 &  1.056$+$01 &  7.985$+$04 &  1.334$-$09 &  1.855$-$10 &  0.000$+$00 &  0.000$+$00 &  6.440$+$01 \\       
    1 &  249 &  1.056$+$01 &  0.000$+$00 &  0.000$+$00 &  0.000$+$00 &  3.403$+$06 &  4.015$+$03 &  0.000$+$00 \\       
    1 &  251 &  1.056$+$01 &  0.000$+$00 &  0.000$+$00 &  0.000$+$00 &  0.000$+$00 &  0.000$+$00 &  7.641$+$00 \\       
    1 &  253 &  1.051$+$01 &  4.325$+$06 &  3.578$-$08 &  4.951$-$09 &  0.000$+$00 &  0.000$+$00 &  2.231$+$02 \\
\hline                                                                                                                                                          
\end{tabular}                                                                                                                                                   
\end{table*}                                                                                                                                                    
\newpage                                                                                                                                                        
\clearpage                                                                                                                                                      
\setcounter{table}{2}                                                                                                                                           
\begin{table*}                                                                                                                                                  
\caption{Transition wavelengths ($\lambda_{ij}$ in ${\rm \AA}$), radiative rates (A$_{ji}$ in s$^{-1}$), oscillator strengths (f$_{ij}$, dimensionless), and line     
strengths (S, in atomic units) for electric dipole (E1), and A$_{ji}$ for E2, M1, and M2 transitions in W LVIII. ($a{\pm}b \equiv a{\times}$10$^{{\pm}b}$).}    
\begin{tabular}{rrrrrrrrr}                                                                                                                                      
\hline                                                                                                                                                          
\hline                                                                                                                                                          
$i$ & $j$ & $\lambda_{ij}$ & A$^{{\rm E1}}_{ji}$  & f$^{{\rm E1}}_{ij}$ & S$^{{\rm E1}}$ & A$^{{\rm E2}}_{ji}$  & A$^{{\rm M1}}_{ji}$ & A$^{{\rm M2}}_{ji}$ \\  
\hline                                                                                                                                                          
    1 &  309 &  9.962$+$00 &  3.454$+$09 &  7.708$-$05 &  1.011$-$05 &  0.000$+$00 &  0.000$+$00 &  4.329$+$03 \\       
    1 &  311 &  9.949$+$00 &  2.511$+$08 &  1.863$-$06 &  2.441$-$07 &  0.000$+$00 &  0.000$+$00 &  1.522$+$02 \\       
    1 &  313 &  9.941$+$00 &  1.242$+$08 &  2.760$-$06 &  3.613$-$07 &  0.000$+$00 &  0.000$+$00 &  2.891$+$01 \\       
    1 &  315 &  9.920$+$00 &  3.129$+$07 &  4.616$-$07 &  6.030$-$08 &  0.000$+$00 &  0.000$+$00 &  7.674$+$01 \\       
    1 &  316 &  9.910$+$00 &  4.007$+$07 &  8.848$-$07 &  1.155$-$07 &  0.000$+$00 &  0.000$+$00 &  1.616$+$02 \\       
    1 &  317 &  9.888$+$00 &  3.693$+$08 &  2.707$-$06 &  3.524$-$07 &  0.000$+$00 &  0.000$+$00 &  1.998$+$02 \\       
    1 &  318 &  9.902$+$00 &  9.157$+$08 &  1.346$-$05 &  1.755$-$06 &  0.000$+$00 &  0.000$+$00 &  1.064$+$03 \\
\hline                                                                                                                                                          
\end{tabular}                                                                                                                                                   
\end{table*}                                                                                                                                                    
\newpage                                                                                                                                                        
\clearpage                                                                                                                                                      
\setcounter{table}{2}                                                                                                                                           
\begin{table*}                                                                                                                                                  
\caption{Transition wavelengths ($\lambda_{ij}$ in ${\rm \AA}$), radiative rates (A$_{ji}$ in s$^{-1}$), oscillator strengths (f$_{ij}$, dimensionless), and line     
strengths (S, in atomic units) for electric dipole (E1), and A$_{ji}$ for E2, M1, and M2 transitions in W LVIII. ($a{\pm}b \equiv a{\times}$10$^{{\pm}b}$).}    
\begin{tabular}{rrrrrrrrr}                                                                                                                                      
\hline                                                                                                                                                          
\hline                                                                                                                                                          
$i$ & $j$ & $\lambda_{ij}$ & A$^{{\rm E1}}_{ji}$  & f$^{{\rm E1}}_{ij}$ & S$^{{\rm E1}}$ & A$^{{\rm E2}}_{ji}$  & A$^{{\rm M1}}_{ji}$ & A$^{{\rm M2}}_{ji}$ \\  
\hline                                                                                                                                                          
    1 &  378 &  9.496$+$00 &  0.000$+$00 &  0.000$+$00 &  0.000$+$00 &  6.038$+$03 &  5.464$+$04 &  0.000$+$00 \\       
    1 &  379 &  9.495$+$00 &  0.000$+$00 &  0.000$+$00 &  0.000$+$00 &  0.000$+$00 &  0.000$+$00 &  8.014$+$02 \\       
    1 &  380 &  9.483$+$00 &  1.264$+$08 &  8.522$-$07 &  1.064$-$07 &  0.000$+$00 &  0.000$+$00 &  1.310$+$02 \\       
    1 &  382 &  9.474$+$00 &  3.308$+$05 &  4.452$-$09 &  5.554$-$10 &  0.000$+$00 &  0.000$+$00 &  1.309$+$01 \\       
    1 &  385 &  9.467$+$00 &  0.000$+$00 &  0.000$+$00 &  0.000$+$00 &  2.151$+$04 &  9.692$+$04 &  0.000$+$00 \\       
    1 &  386 &  9.461$+$00 &  4.538$+$08 &  9.134$-$06 &  1.138$-$06 &  0.000$+$00 &  0.000$+$00 &  6.463$+$01 \\       
    1 &  387 &  9.427$+$00 &  0.000$+$00 &  0.000$+$00 &  0.000$+$00 &  1.288$+$03 &  1.039$+$02 &  0.000$+$00 \\
\hline                                                                                                                                                          
\end{tabular}                                                                                                                                                   
\end{table*}        

\end{document}